\documentstyle[aps]{revtex}
\def\be{\begin{equation}}
\def\ee{\end{equation}}
\def\l{\left}
\def\r{\right}
\def\ba{\begin{array}}
\def\ea{\end{array}}
\def\bea{\begin{eqnarray}}
\def\eea{\end{eqnarray}}

\def\p{\partial}
\def\f{\frac}
\def\tr{{\rm tr}}

\begin{document}

\setcounter{equation}{0}
\renewcommand{\theequation}{\arabic{section}.\arabic{equation}}

\markboth{M. Kubo, Z. Maki, M. Nakahara and T. Saito}
{Grand Unification from Gauge Theory on $M_4 \times Z_N$}

\title{Grand Unification from Gauge Theory on $M_4 \times Z_N$}
\author{%       %Use \sc for the family name
Masahiro Kubo\footnote{fwiw9993@mb.infoweb.ne.jp},
Ziro Maki\footnote{ziromaki@phys.kindai.ac.jp},
Mikio Nakahara\footnote{nakahara@math.kindai.ac.jp} and
Takesi Saito$^{\dagger}$\footnote{tsaito@jpnyitp.yukawa.kyoto-u.ac.jp}}
\address{Department of Physics, Kinki University, Higashi-Osaka 577-8502
\\
$^{\dagger}$Department of Physics, Kwansei Gakuin University,
Nishinomiya 662-0891}
\date{To be published in Prog. Theor. Phys. {\bf 100} (1998) No. 1}

\maketitle
\begin{abstract}
The SU(5) grand unified theory (GUT) is derived from the geometrical point 
of view of gauge theory on three-sheeted space-time,i.e., $M_4
\times Z_3$ manifold, 
which was previously proposed in Ref. 4 without recourse to noncommutative 
geometry. A derivation of SO(10) GUT
is also discussed in the same point of view.
\end{abstract}
%
%\newpage
\section{Introduction}

Recently it has become clear that spontaneously broken gauge theories 
based on the Higgs mechanism are all gauge theories on
$M_4 \times Z_N$, where $M_4$
is the four-dimensional Minkowski space and $Z_N$ is the discrete space 
with $N$ points. \cite{ref:1} \cite{ref:2}
In these theories the Higgs fields are regarded as 
gauge fields along directions in the discrete space. The bosonic part of the 
action is made of the pure Yang--Mills actions containing gauge fields in both 
continuous and discrete spaces, and the Yukawa couplings are just
gauge interactions of fermions.

The Weinberg--Salam (WS) \cite{ref:3}
model for electroweak interactions is also one 
of such gauge theories with the manifold $M_4 \times Z_2$.
In this case the most striking result is that the Higgs potential
takes the form which necessarily leads to spontaneous gauge symmetry
breaking. This is due to the fact that the Higgs potential is 
given by 
\begin{equation}
V(H) = \lambda \left(H^{\dagger} H-1\right)^2,\ \ (\lambda > 0)
\end{equation}
where $H$ is the Higgs field and $H^{\dagger}H-1$ corresponds to the
curvature along $Z_2$. Thus we have a possible explanation of the
geometrical origin of the Higgs field
and the symmetry breaking mechanism, which has puzzled physicist for 
many years.

At first, gauge theories on $M_4 \times Z_N$
have been formulated in terms of noncommutative geometry (NCG)
by Connes\cite{ref:1} and its various alternative
versions.\cite{ref:2}
Any approach based on NCG, however, has been too algebraic so far,
rather than geometric, and hence
its geometrical meaning was not very clear. Recently one of the
authors (T.S.), in collaboration with Konisi,\cite{ref:4} has constructed
another gauge theory on $M_4 \times Z_N$
from purely geometrical point of view, without employing entire context
of NCG.
In this approach the Higgs fields have been introduced as mapping 
functions between any pair of vector fields each of which is defined on
a sheet in the $N$-sheeted space-time independently.
This theory has been applied to WS 
model, $N=2$ and $4$ super Yang--Mills theories and the Brans--Dicke theory
of gravity, and geometric structures of them have been clarified.\cite{ref:5}

The purpose of the present paper is to apply this `geometric' formalism,
in contrast with algebraic one, to SU(5) and SO(10) GUTs.\cite{ref:6}
The geometric structure of
the theories become manifest in our approach compared to other works based 
on NCG.\cite{ref:7} The SU(5) GUT in its simplest form
is ruled out as a true theory
because, among others, its prediction of proton lifetime is much shorter
than the observed lower bound and the predicted Weinberg angle 
does not agree with accurate experiments by the LEP.
However, the reconstruction of SU(5) in our framework
is still important since this is a typical pattern of GUTs. 
Furthermore, this is also helpful in constructing more
realistic models such as SO(10) GUT.

In SU(5) GUT we need two kinds of Higgs fields, $H_5$ and $\phi_{24}$,
belonging to
5- and 24-dimensional representations of SU(5), respectively. From this
reason we should prepare at least 3-sheeted space-time, i.e., $M_4 \times
Z_3$ as the manifold.
The Higgs potential will be constructed from curvatures in this $Z_3$ 
space. Contrary to $M_4 \times Z_2$
case, it is not so evident whether this potential gives rise to
spontaneous gauge symmetry breaking.
%This important problem will be discussed in detail. We will find that
%$\phi_{24}$ can certainly have non-zero vacuum expectation value if this
%potential does not contain a term linear in $\phi_{24}$.

In Section 2 we consider SU(5) gauge fields on $M_4 \times Z_3$
and construct the Higgs potential.
In Section 3 we give Lagrangian for fermionic fields on the same
manifold. The final section is devoted to concluding remarks including SO(10) 
GUT construction.

\section{Bosonic sector of SU(5) GUT}
\setcounter{equation}{0}

Let us consider a manifold $M_4 \times Z_3$, where $M_4$
is the 4-dimensional space-time and $Z_3$ the discrete space with three
points $g_p\ (p=0,1,2)$. The latter points are subject to
the algebraic relations
\begin{equation}
\begin{array}{c}
g_0+g_0=g_0,\ g_0+g_1=g_1,\ g_0+g_2=g_2,\\
g_1+g_1=g_2,\ g_1+g_2=g_0,\ g_2+g_2=g_1,
\end{array}
\end{equation}
where additive notation is used for the group product.
Let us attach a complex 5-dimensional internal vector space $V[5,x,g_p]$
to every point $(x, g_p) \in M_4 \times Z_3$.
The fermionic fields $\psi(x, g_p)$ are chosen as
\begin{equation}
\psi(x, g_0)=\psi_5(x) = \left(
\begin{array}{c}
d^1\\
d^2\\
d^3\\
e^+\\
\bar{\nu}_e
\end{array}
\right)_R,
\end{equation}
and
\begin{equation}
\psi(x, g_1)=\psi(x, g_2) = \psi_{10}(x)
= \left(
\begin{array}{ccccc}
0&u^c_3&-u^c_2&-u^1&-d^1\\
-u^c_3&0&u^c_1&-u^2&-d^2\\
u^c_2&-u^c_1&0&-u^3&-d^3\\
u^1&u^2&u^3&0&-e^+\\
d^1&d^2&d^3&e^+&0
\end{array}
\right)_L,
\end{equation}
where $\psi_5$ and $\psi_{10}$ stand for the 5-dimensional fundamental
representation and the antisymmetric 10-dimensional representation of SU(5)
(see Fig. 1).\footnote{Alternatively, one could employ the assignments
$\psi(x, g_0) = \psi_{10}(x)$ and $\psi(x, g_1) = \psi(x, g_2) = \psi_5(x)$,
which lead to essentially the same Lagrangian as the present one.
Details will be reported in a subsequent publication.}
\begin{figure}
\setlength{\unitlength}{1mm}
\begin{center}
\begin{picture}(100,40)(0,5)
\put(35,35){\line(1,0){30}}
\put(35,35){\line(2,-3){15}}
\put(50,12.5){\line(2,3){15}}
\put(30,32){$g_1$}
\put(67,32){$g_2$}
\put(48,7.5){$g_0$}
\put(30,38){$\psi_{10}(x)$}
\put(60,38){$\psi_{10}(x)$}
\put(53,12.5){$\psi_5(x)$}
\end{picture}
\end{center}
\caption{Assignment of fermion fields on $Z_3$.}
\label{fig.1}
\end{figure}
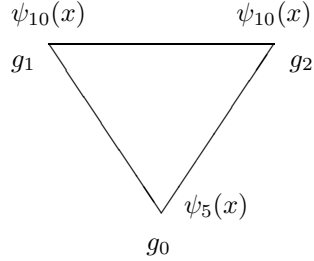

The conventional SU(5) Yang--Mills field $A_{\mu}(x)$ coupled to $\psi_5$
and $\psi_{10}$ are introduced through covariant derivatives
\begin{eqnarray}
\left(D_{\mu} \psi_5(x)\right)^i&=&\partial_{\mu}\psi^i_5(x)
-ig(A_{\mu}(x))^i_{\ j} \psi_5^j(x),\\
\left(D_{\mu} \psi_{10}(x)\right)^{ij}&=&
\partial_{\mu} \psi_{10}^{ij}(x)-i\frac{1}{2}g(A_{\mu}(x))^i_{\ k}
\psi_{10}^{kj}(x)+i\frac{1}{2}g(A_{\mu}(x))^j_{\ k}
\psi_{10}^{ki}(x),\nonumber\\
&=& \f{1}{2}\l[\p_{\mu} \Delta^{ij}_{lm}-i\f{1}{2}g (A_{\mu})^{i}_{\ k}
\Delta^{kj}_{lm} + i\f{1}{2}g (A_{\mu})^{j}_{\ k} \Delta^{ki}_{lm}\r]
\psi_{10}^{lm}(x),
\end{eqnarray}
where
\begin{eqnarray}
A_{\mu}(x)&=&A_{\mu}^a(x)T_a,\ {\rm tr}(T_aT_b)=\delta_{ab}/2,\ (a=1,2,\cdots
24)\nonumber\\
\Delta_{lm}^{ij}&=&\delta_l^i \delta_m^j - \delta_l^j \delta_m^i
= - \Delta_{ml}^{ij}.
\end{eqnarray}
and $T_a$ is the generator of SU(5). The curvature (or field strength) 
of $A_{\mu}(x)$ is given by 
\begin{equation}
F_{\mu\nu}(x)=\partial_{\mu} A_{\nu}-\partial_{\nu} A_{\mu}
-i g[A_{\mu}, A_{\nu}].
\end{equation}

Let us consider a mapping of $\psi(x, g_0)$ from $(x, g_0)$ to
$(x+\delta x, g_0)$. The mapped vector $\psi_{\parallel}(x+\delta x, g_0)$
is given by 
\be
\psi_{\parallel}^i(x+\delta x, g_0)
=H^i_{\ j}(x+\delta x, x, g_0) \psi^j(x,g_0),
\ee
where the mapping function $H^i_{\ j}(x+\delta x, x, g_0)$ is expanded as 
\be
H^i_{\ j}(x+\delta x, x, g_0) = \l(1+i g A_{\mu}(x) \delta x^{\mu}
+ \cdots \r)^i_{\ j},
\ee
and $A_{\mu}(x)$ is identified with the above SU(5) Yang--Mills field.
The mapping function obeys the transformation rule
\be
H(x+\delta x, x, g_0) \to \tilde{H}(x+\delta x, x, g_0)=
U(x+\delta x, g_0) H(x+\delta x, x, g_0) U^{-1}(x, g_0)
\ee
under an SU(5) gauge transformation
\be
\psi(x, g_0) \to \tilde{\psi}(x, g_0) = U(x, g_0) \psi(x, g_0),
\ee
where $U(x, g_0)$ is parametrized by an arbitrary function
$\theta^a(x, g_0)$ as
\be
U(x, g_0) = \exp\l\{i\theta^a(x, g_0) T_a \r\}. 
\ee
The gauge transformation of the form (2.10) has been employed
in the lattice gauge theories and is reduced to the familiar form
\be
A_{\mu}(x) \to \tilde{A}_{\mu}(x) = U(x, g_0) A_{\mu}(x) U^{-1}(x, g_0)
-i \f{1}{g} \p_{\mu} U(x, g_0) U^{-1}(x, g_0).
\ee
From this reason we sometimes call $H(x+\delta x, x, g_0)$
itself the gauge field.

As is well known, the curvature $F_{\mu\nu}$ associated with the
connection $A_{\mu}$ has a geometircal interpretation as
a difference between two images (or parallel transportations)
of a vector $\psi(x,g_0)$ from $x$ 
to $x+\delta_1 x + \delta_2 x$ along two paths $C_1$ and
$C_2$ depicted in Fig.2. 
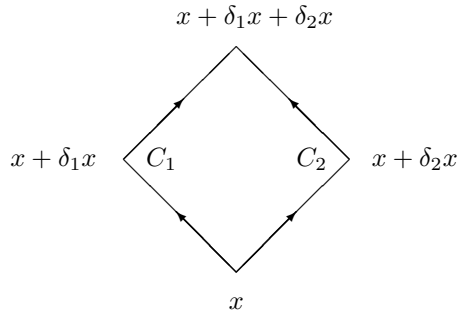
\begin{figure}
\setlength{\unitlength}{1mm}
\begin{center}
\begin{picture}(100,45)(0,0)
\put(35,25){\line(1,1){15}}
\put(35,25){\line(1,-1){15}}
\put(65,25){\line(-1,1){15}}
\put(65,25){\line(-1,-1){15}}
\put(35,25){\vector(1,1){8}}
\put(50,10){\vector(-1,1){8}}
\put(65,25){\vector(-1,1){8}}
\put(50,10){\vector(1,1){8}}
\put(42,43){$x+\delta_1 x+\delta_2 x$}
\put(20,24){$x+\delta_1 x$}
\put(68,24){$x+\delta_2 x$}
\put(49,5){$x$}
\put(38,24){$C_1$}
\put(58,24){$C_2$}
\end{picture}
\end{center}
\caption{Two paths from which the curvature $F_{\mu \nu}$ arises.}
\label{fig.2}
\end{figure}
In the same way  
we consider the curvature $F_{\mu g_1}(x, g_0)$ defined by Fig.3, which 
arises from the difference between two images of $\psi(x, g_0)$
from $(x, g_0)$ to $(x+\delta x, g_1)$ along two paths $C_3$ and $C_4$.
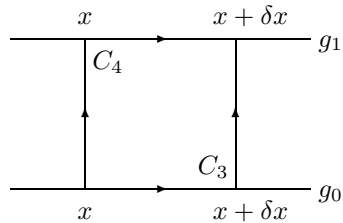
\begin{figure}
\setlength{\unitlength}{1mm}
\begin{center}
\begin{picture}(100,40)(0,0)
\put(30,10){\line(1,0){40}}
\put(30,30){\line(1,0){40}}
\put(40,10){\line(0,1){20}}
\put(60,10){\line(0,1){20}}
\put(40,30){\vector(1,0){11}}
\put(40,10){\vector(1,0){11}}
\put(40,10){\vector(0,1){11}}
\put(60,10){\vector(0,1){11}}
\put(39,32){$x$}
\put(39,6){$x$}
\put(57,32){$x+\delta x$}
\put(57,6){$x+\delta x$}
\put(41,26){$C_4$}
\put(55,12){$C_3$}
\put(71,9){$g_0$}
\put(71,29){$g_1$}
\end{picture}
\end{center}
\caption{Two paths which define the curvature $F_{\mu g_0}$.}
\label{fig.3}
\end{figure}
In order to calculate this curvature, we introduce first a 
mapping of $\psi(x, g_0)$ from $(x, g_0)$ to $(x, g_1)$. 
The mapped function $\psi_{\parallel}^{ij}(x, g_1)$ should be 
antisymmetric with respect to $i$ and $j$, and is given by
\be
\psi_{\parallel}^{ij}(x, g_1) = H^{ij}_{\ \ l}(x, g_1, g_0)
\psi^l(x, g_0),
\ee
where the mapping function $H^{ij}_{\ \ l}(x, g_1, g_0)$ is expressed in
terms of $H_5^i(x)$, the Higgs 
field in the 5-dimensional fundamental representation of SU(5), as
{
\setcounter{enumi}{\value{equation}}
\addtocounter{enumi}{1}
\setcounter{equation}{0}
\renewcommand{\theequation}{\arabic{section}.\theenumi\alph{equation}}
\be
H^{ij}_{\ \ l}(x, g_1, g_0)=\f{1}{2}\l[H^i(x) \delta^j_l
-H^j(x) \delta^i_l \r],\ \ \l(H^i \equiv H^i_5\r)
\ee
or, in a tensor product form, as
\be
H(x, g_1, g_0)=\f{1}{2} \l[H(x) \otimes 1-1 \otimes H(x)\r].
\ee
\setcounter{equation}{\value{enumi}}
}
Under the gauge transformations at $(x, g_0)$ and $(x, g_1)$
it obeys the rule
\be
H(x, g_1, g_0) \to \tilde{H}(x, g_1, g_0)=U(x, g_1) \otimes
U(x, g_1) H(x, g_1, g_0) U^{-1}(x, g_0).
\ee
This has essentially the same form as that of (2.10). Therefore, we
regard $H(x, g_1, g_0)$ as a gauge field along $g_0-g_1$ direction.
Note that the mapping function (2.15a) cannot be unitary (see Appendix).

Now, let us calculate the curvature $F_{\mu g_1}(x, g_0)$.
The two mappings are given by
\bea
\psi^{ij}_{C_3} &=&
H^{ij}_{\ \ k}(x+\delta x, g_1, g_0)H^{k}_{\ l}(x+\delta x, x, g_0)
\psi^{l}(x, g_0),\\
\psi^{ij}_{C_4} &=&H^{ij}_{\ \ km}(x+\delta x, x, g_1)
H^{km}_{\ \ l}(x, g_1, g_0)\psi^{l}(x, g_0),
\end{eqnarray}
where $H^k_{\ l}$ and $H^{ij}_{\ \ k}$ are defined by Eqs. (2.9) and
(2.15a), respectively, and
\begin{eqnarray}
H^{ij}_{\ \ km}(x+\delta x, x, g_1)&=&\frac{1}{2}\left[
\Delta^{ij}_{km} + i\f{1}{2} g\l((A_{\mu}(x))^i_{\ l}\Delta^{lj}_{km}
-(A_{\mu}(x))^j_{\ l}\Delta^{li}_{km}\r)\delta x^{\mu} \r]\nonumber\\
&=& H^{ij}_{\ \ km}(x+\delta x, x, g_2).
\end{eqnarray}
Taking the difference between them, we have
{
\setcounter{enumi}{\value{equation}}
\addtocounter{enumi}{1}
\setcounter{equation}{0}
\renewcommand{\theequation}{\arabic{section}.\theenumi\alph{equation}}
\be
\psi^{ij}_{C_3}-\psi^{ij}_{C_4}=(F_{\mu g_1}(x, g_0))^{ij}_{\ \ l}
\delta x^{\mu} \psi^l(x, g_0),
\ee
where
\bea
(F_{\mu g_1}(x, g_0))^{ij}_{\ \ l}&=&
\f{1}{2}\l[\p_{\mu} H^i \delta^j_l -\f{1}{2} ig\l((A_{\mu})^i_{\ k}
H^k
\delta^j_l-H^i (A_{\mu})^j_{\ l}\r) - (i \leftrightarrow j)\r]\nonumber\\
&=& \f{1}{2}\l[\l(D_{\mu} H\r)^{ij}_{\ \ l}-(i \leftrightarrow j)\r]\nonumber\\
&\equiv& {\l(D_{\mu} H\r)^{[ij]}}_{l}.
\eea

The Yang--Mills action associated with this curvature is
just the kinetic term for a Higgs field $H_5$. In the same way 
we find
\bea
{(F_{\mu g_0}(x, g_1))}^{ij}_{\ \ l}
&=&\l(D_{\mu} H^{\dagger} \r)^{[ij]}_{\ \ \ l},\\
{(F_{\mu g_2}(x, g_0))}^{ij}_{\ \ l}&=&\l(D_{\mu} H\r)^{[ij]}_{\ \ l},\\
{(F_{\mu g_0}(x, g_2))}^{ij}_{\ \ \ l}
&=&\l(D_{\mu} H^{\dagger} \r)^{[ij]}_{\ \ \ l}.
\eea
\setcounter{equation}{\value{enumi}}
}

Another curvature $F_{\mu g_2}(x, g_1)$ may be defined by Fig.4.
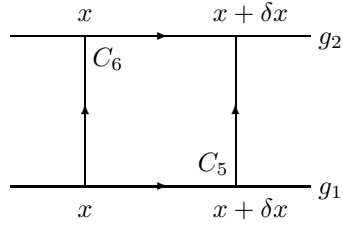
\begin{figure}
\setlength{\unitlength}{1mm}
\begin{center}
\begin{picture}(100,40)(0,0)
\put(30,10){\line(1,0){40}}
\put(30,30){\line(1,0){40}}
\put(40,10){\line(0,1){20}}
\put(60,10){\line(0,1){20}}
\put(40,30){\vector(1,0){11}}
\put(40,10){\vector(1,0){11}}
\put(40,10){\vector(0,1){11}}
\put(60,10){\vector(0,1){11}}
\put(39,32){$x$}
\put(39,6){$x$}
\put(57,32){$x+\delta x$}
\put(57,6){$x+\delta x$}
\put(41,26){$C_6$}
\put(55,12){$C_5$}
\put(71,9){$g_1$}
\put(71,29){$g_2$}
\end{picture}
\end{center}
\caption{Two paths $C_5$ and $C_6$ by which the curvature
$F_{\mu g_2}(x, g_1)$ is defined.}
\label{fig.4}
\end{figure}
The mapping of $\psi^{ij}(x, g_1)$ from $(x, g_1)$ to $(x, g_2)$
is defined by the traceless Higgs 
field $\phi_{24}$ of the 24-dimensional representation, i.e.,
\be
\psi^{ij}_{\|}(x, g_2) = H^{ij}_{\ \ kl}(x, g_2, g_1)
\psi^{kl}(x, g_1),
\ee
where
\be
H^{ij}_{\ \ kl}(x, g_2, g_1)=\f{1}{4} \l[\phi^i_{m}(x) \Delta^{mj}_{kl}
- \phi^j_{m}(x) \Delta^{mi}_{kl}\r]
\ee
with $\phi^i_{k} \equiv \l(\phi_{24}\r)^i_{k}$ and
$\tr\,\phi_{24} = 0$.
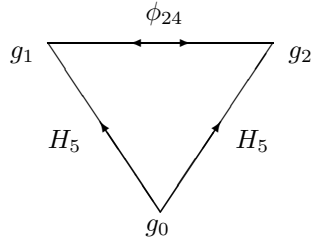
\begin{figure}
\setlength{\unitlength}{1mm}
\begin{center}
\begin{picture}(100,40)(0,5)
\put(35,35){\line(1,0){30}}
\put(35,35){\line(2,-3){15}}
\put(50,12.5){\line(2,3){15}}
\put(30,33){$g_1$}
\put(67,33){$g_2$}
\put(48,10){$g_0$}
\put(50,12.5){\vector(-2,3){8}}
\put(50,12.5){\vector(2,3){8}}
\put(50,35){\vector(1,0){4}}
\put(50,35){\vector(-1,0){4}}
\put(48,38){$\phi_{24}$}
\put(35,21){$H_5$}
\put(60,21){$H_5$}
\end{picture}
\end{center}
\caption{The mapping function ${H^{ij}}_l(x, g_1, g_0)$ from $(x, g_0)$ to
$(x, g_1)$ is expressed in terms of the Higgs field $H_5(x)$. The same is
true for ${H^{ij}}_l(x, g_2, g_0)$. The mapping function ${H^{ij}}_{kl}
(x, g_2, g_1)$ between $(x, g_1)$ and $(x, g_2)$ is expressed in terms of
the Higgs field $\phi_{24}(x)$.}
\label{fig.5}
\end{figure}
Therefore, two mappings of $\psi^{ij}(x, g_1)$ from $(x, g_1)$ to
$(x+\delta x, g_2)$ along paths $C_5$
and $C_6$ in Fig.4 are given by
\bea
\psi^{ij}_{C_5}
%(x+\delta x, g_2)
&=& {H^{ij}}_{mn}(x+\delta x, g_2, g_1){H^{mn}}_{kl}(x+\delta x, x, g_1)
\psi^{kl}(x, g_1)\\
\psi^{ij}_{C_6}&=&H^{ij}_{\ \ mn}(x+\delta x, x, g_2)
{H^{mn}}_{kl}(x, g_2, g_1)\psi^{kl}(x, g_1).
\eea
Substituting Eqs. (2.19) and (2.22) into Eqs. (2.23) and (2.24),
we have the curvature $F_{\mu g_2}(x, g_1)$, i.e.,
\be
\psi^{ij}_{C_5}-\psi^{ij}_{C_6} =
{(F_{\mu g_2}(x, g_1))}^{ij}_{\ \ kl} \delta x^{\mu} \psi^{kl}(x, g_1),
\ee
where
\bea
{(F_{\mu g_2}(x, g_1))}^{ij}_{\ \ kl} &=& \f{1}{4} 
\l[\l(\p_{\mu} \phi-ig[A_{\mu}, \phi]\r)^i_m \Delta^{mj}_{kl}
  -\l(\p_{\mu} \phi-ig[A_{\mu}, \phi]\r)^j_m \Delta^{mi}_{kl}\r]\nonumber\\
&\equiv& \l(D_{\mu} \phi \r)^{[ij]}_{\ \ \ [kl]}\nonumber\\
&=& {(F_{\mu g_1}(x, g_2))}^{ij}_{\ \ kl}
\eea

There are also six kinds of curvatures defined on the discrete space.
The first kind, defined by Fig. 6(a), is the same as that in the WS case.
\begin{figure}
\setlength{\unitlength}{1mm}
\begin{center}
\begin{picture}(150,20)(0,0)
\put(10,8){\line(1,0){25}}
\put(10,12){\line(1,0){25}}
\put(35,10){\oval(4,4)[r]}
\put(27,8){\vector(-1,0){5}}
\put(20,12){\vector(1,0){5}}
\put(9,10){\circle*{2}}
\put(37,10){\circle*{2}}
\put(4,9){$g_1$}
\put(39,9){$g_2$}
\put(22,14){$C_7$}
\put(22,3){(a)}
\put(60,8){\line(1,0){25}}
\put(60,12){\line(1,0){25}}
\put(85,10){\oval(4,4)[r]}
\put(77,8){\vector(-1,0){5}}
\put(70,12){\vector(1,0){5}}
\put(59,10){\circle*{2}}
\put(87,10){\circle*{2}}
\put(54,9){$g_0$}
\put(89,9){$g_1$}
\put(72,14){$C_8$}
\put(72,3){(b)}
\put(110,8){\line(1,0){25}}
\put(110,12){\line(1,0){25}}
\put(110,10){\oval(4,4)[l]}
\put(127,12){\vector(-1,0){5}}
\put(120,8){\vector(1,0){5}}
\put(108,10){\circle*{2}}
\put(136,10){\circle*{2}}
\put(103,9){$g_0$}
\put(138,9){$g_1$}
\put(122,14){$C_9$}
\put(122,3){(c)}
\end{picture}
\end{center}
\caption{The paths defining curvatures of (a) the first, (b) the second
and (c) the third kinds.}
\label{fig.6}
\end{figure}
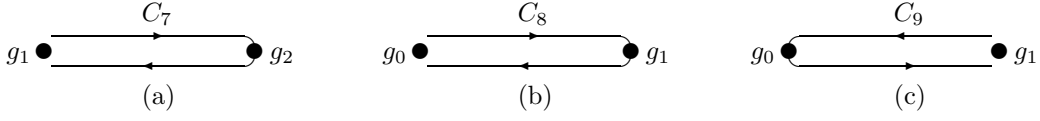
That is, $\psi^{ij}(x, g_1)$ is compared with ${\psi}^{ij}_{C_7}(x, g_1)$
which is obtained by mapping $\psi(x, g_1)$ from $(x, g_1)$ to $(x, g_2)$
and then back to $(x, g_1)$ through the path $C_7$, i.e.,
\bea
\lefteqn{{\psi}^{ij}_{C_7}(x, g_1)-\psi^{ij}(x, g_1)}\nonumber\\
&=& H^{ij}_{\ \ mn}(x, g_1, g_2)
H^{mn}_{\ \ \ \ kl}(x, g_2, g_1)\psi^{kl}(x, g_1)-\f{1}{2} \Delta^{ij}_{kl}
\psi^{kl}(x, g_1)\nonumber\\
&=& (F_{(1)})^{ij}_{\ \ kl} \psi^{kl}(x, g_1).
\eea
Substituting Eqs. (2.21b) and (2.6) into (2.27),
we get the first kind of curvature
\bea
(F_{(1)})^{ij}_{\ \ kl}
&=&\f{1}{4} \l(\phi^i_m \delta^j_n - \phi^j_m \delta^i_n\r)
\l(\phi^m_k \delta^n_l - \phi^n_k \delta^m_l \r)-\f{1}{2} \Delta^{ij}_{kl}
\nonumber\\
&=& \f{1}{4} (\phi^i_m \phi^m_k \delta^j_l -
\phi^j_m \phi^m_k \delta^i_l + \phi^j_l \phi^i_k - \phi^i_l \phi^j_k )
- \f{1}{2} \Delta^{ij}_{kl}.
\eea
Note that there is no need to antisymmetrize $F$ with respect to $k$
and $l$, since it will be automatically 
taken into account when constructing the Lagrangian.
In a similar way we find the second and third kinds of 
curvatures for Fig. 6(b) and Fig. 6(c), respectively,
\bea
{(F_{(2)})}^{ij}
&=& \f{1}{2} \l[\l(H^{\dagger}H-2\r)\delta^{ij}-H^{\dagger i} H^{j}
\r]\\
{(F_{(3)})}^{ij}_{\ \ kl}&=& \f{1}{2} \l(H^i H_k^{\dagger} \delta^j_l
-H^j H_k^{\dagger} \delta^i_l - \Delta^{ij}_{kl}\r).
\eea

Other three kinds of curvatures will arise from triangle diagrams
shown in Fig. 7.
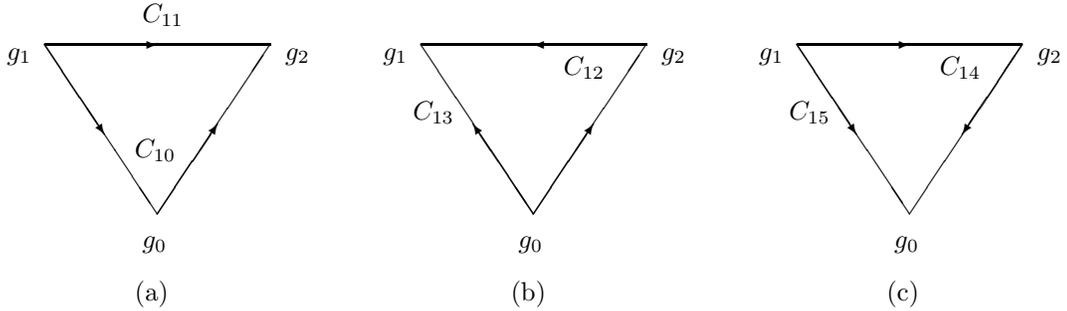
\begin{figure}
\setlength{\unitlength}{1mm}
\begin{center}
\begin{picture}(150,45)(0,0)
\put(10,35){\line(1,0){30}}
\put(10,35){\line(2,-3){15}}
\put(25,12.5){\line(2,3){15}}
\put(5,33){$g_1$}
\put(42,33){$g_2$}
\put(23,8){$g_0$}
\put(10,35){\vector(2,-3){8}}
\put(25,12.5){\vector(2,3){8}}
\put(25,35){\vector(1,0){0}}
\put(23,38){$C_{11}$}
\put(22,20){$C_{10}$}
\put(22,1){(a)}
\put(60,35){\line(1,0){30}}
\put(60,35){\line(2,-3){15}}
\put(75,12.5){\line(2,3){15}}
\put(55,33){$g_1$}
\put(92,33){$g_2$}
\put(73,8){$g_0$}
\put(75,12.5){\vector(-2,3){8}}
\put(75,12.5){\vector(2,3){8}}
\put(75,35){\vector(-1,0){0}}
\put(59,25){$C_{13}$}
\put(79,31){$C_{12}$}
\put(72,1){(b)}
\put(110,35){\line(1,0){30}}
\put(110,35){\line(2,-3){15}}
\put(125,12.5){\line(2,3){15}}
\put(105,33){$g_1$}
\put(142,33){$g_2$}
\put(123,8){$g_0$}
\put(110,35){\vector(2,-3){8}}
\put(140,35){\vector(-2,-3){8}}
\put(125,35){\vector(1,0){0}}
\put(109,25){$C_{15}$}
\put(129,31){$C_{14}$}
\put(122,1){(c)}
\end{picture}
\end{center}
\caption{The paths defining curvatures of (a) the fourth, (b) the fifth
and (c) the sixth kinds.}
\label{fig.7}
\end{figure}
They are given by
\bea
{(F_{(4)})}^{ij}_{\ \ kl}&=&\f{1}{2} \l(H^i H_k^{\dagger}
\delta^j_l-H^j H_k^{\dagger} \delta^i_l
- \phi^i_k \delta^j_l + \phi^j_k \delta^i_l \r),\\
{(F_{(5)})}^{ij}_{\ \ m}
&=&\f{1}{4} \l(\phi^i_k \delta^j_l - \phi^j_k \delta^i_l \r)
\l(H^{k}\delta^l_m - H^{l}\delta^k_m\r)
-\f{1}{2} \l(H^{i} \delta^j_m - H^{j}\delta^i_m \r),\nonumber\\
& &\\
{(F_{(6)})}^j_{\ kl}&=& \f{1}{4}\l(H_i^{\dagger} \phi^i_k \delta^j_l
-H_i^{\dagger} \phi^i_l \delta^j_k + H_k^{\dagger} \phi^j_l
- H_l^{\dagger} \phi^j_k
-2 H_k^{\dagger} \delta^j_l + 2 H_l^{\dagger} \delta^j_k\r),\nonumber\\
& &
\eea
for Figs. 7(a), 7(b) and 7(c), respectively.

Let us then construct a Lagrangian for the bosonic sector.
Since the mapping functions are all gauge fields, the Lagrangian should be of 
the Yang--Mills type, i.e.,
\bea
{\cal L}_B&=& -\f{1}{2} \tr\l(F^{\dagger}_{\mu\nu} F^{\mu\nu}\r)
+ A \tr\l(F^{\dagger}_{\mu g_2}(g_1) F^{\mu g_2}(g_1)\r)\nonumber\\
& & + B\tr\l(F^{\dagger}_{\mu g_1}(g_0) F^{\mu g_1}(g_0)\r)
-V(\phi, H)\nonumber\\
&=& -\f{1}{2} \tr\l(F^{\dagger}_{\mu\nu} F^{\mu\nu}\r) 
+ A \l(D^{\mu} \phi \r)^{\dagger\ [kl]}_{\ \ [ij]}
\l(D_{\mu} \phi \r)^{\ [ij]}_{\ \ [kl]}\nonumber\\
& &+ B\l(D^{\mu} H\r)^{\dagger\ l}_{\ \ [ij]} \l(D_{\mu} H\r)^{[ij]}_{\ \ l}
-V(\phi, H),
\eea
where
\bea
V(\phi, H)&=& a\ \tr\l(F_{(1)}^{\dagger} F_{(1)}\r) 
+ b\ \tr\l(F_{(2)}^{\dagger} F_{(2)}\r)
+ c\ \tr\l(F_{(4)}^{\dagger} F_{(4)} \r) 
+ d\ \tr \l(F_{(5)}^{\dagger} F_{(5)}\r)\nonumber\\
&=& a\l[\l(\tr \phi^2\r)^2-\tr \phi^4-8 \tr \phi^2\r]
+ b\l(H^{\dagger} H -2 \r)^2 \nonumber\\
& & + c\l[3 \tr \phi^2 -6 H^{\dagger} \phi H
+ 4 \l(H^{\dagger} H\r)^2 \r]\nonumber\\
& & + d\l[H^{\dagger} \phi \phi H + \tr \phi^2 \l(H^{\dagger} H\r)
-12 H^{\dagger} \phi H+ 16 H^{\dagger} H\r].
\eea
Here the parameters $A$ and $B$ are all real positive so that the Higgs 
kinetic terms are positive. The other parameters $a, b, c$ and $d$
are also real and positive 
so that each of $\tr \l(F_{(i)}^{\dagger} F_{(i)}\r)$, $(i=1, 2, 4, 5)$
is positive for large values of $|\phi|$ and $|H|$. In Eqs. (2.34)
and (2.35) we have used the facts that the Higgs kinetic terms are invariant
under the exchange between 1 and 2, and $\tr\l(F_{(2)}^{\dagger} F_{(2)}\r)=
\tr \l(F_{(3)}^{\dagger} F_{(3)}\r)$ and $\tr \l(F_{(5)}^{\dagger} F_{(5)}\r)
=\tr \l(F_{(6)}^{\dagger} F_{(6)}\r)$. 
The term $V(\phi, H)$ is just the Higgs potential.
In addition to Eq. (2.35) there is another possible 
term $H^{\dagger} \phi H$ which arises from a scalar curvature
that is defined by the difference between a mapping of
$\psi(x, g_0)$ along a closed path $g_0 \to g_1 \to g_2 \to g_0$
and $\psi(x, g_0)$ itself, and 
is the same for cyclic permutations. This is not of the Yang--Mills type,
but is renormalizable, and hence we can add the term $e H^{\dagger} \phi H$,
$e$ being a real parameter, to Eq. (2.35).
Thus we have finally obtained the Higgs potential
\bea
V_f(\phi, H)&=& a \l[ \l(\tr \phi^2\r)^2 - \tr \phi^4\r]-
(8a - 3c) \tr \phi^2 + (b+4c)\l(H^{\dagger} H\r)^2\nonumber\\
& &-4(b-4d)\l(H^{\dagger} H\r)+d\l[H^{\dagger} \phi \phi H + \tr \phi^2
\l(H^{\dagger} H\r)\r]\nonumber\\
& & +(-6c-12 d+e)H^{\dagger} \phi H.
\eea

This should be compared with the conventional phenomenological Higgs 
potential
\bea
V_c(\phi, H)&=& -\f{1}{2} \mu^2 \tr \phi^2 + \f{1}{4} a \l( \tr \phi^2\r)^2
+ \f{1}{4} b\ \tr  \phi^4- \f{1}{2} v^2 H^{\dagger} H\nonumber\\
& &+ \f{1}{4} \lambda \l(H^{\dagger} H\r)^2
+ \alpha\ \tr \phi^2\l(H^{\dagger} H\r)+\beta\ H^{\dagger} \phi^2 H,
\eea
which contains seven parameters, whereas our potential (2.36) contains six 
parameters and the Yukawa coupling term $H^{\dagger} \phi H$
in the tree level. Gauge symmetries are spontaneously broken, in our model, 
when $a$ and $b$
are sufficiently large compared with $c$ and $d$, respectively.

\section{Fermionic sector of SU(5) GUT}
\setcounter{equation}{0}

The Dirac Lagrangian $i \bar{\psi} \gamma^{\mu}D_{\mu} \psi$ with
$D_{\mu} = \p_{\mu} - ig A_{\mu}(x)$, on $M_4$ will be extended to that 
on $M_4 \times Z_3$. This is given by
\be
{\cal L}_F = \sum_{p=0}^2 i \bar{\psi}(x, g_p)\l[\gamma^{\mu} \tau_0
D_{\mu}(p) + \kappa \gamma_5 \l(\tau_1 D_1 + \tau_2 D_2\r)\r]
\psi(x, g_p),
\ee
where $\kappa$ is a mass parameter, $\tau_1$ and $\tau_2$ are Pauli matrices,
$\tau_0$ the $2 \times 2$ unit matrix, and
\be
\psi(x, g_0) = \l(
\ba{c}
0\\
\psi_5(x)
\ea
\r),\ \psi(x, g_1) = \psi(x, g_2)=\l(
\ba{c}
\psi_{10}(x)/\sqrt{2}\\
0
\ea
\r).
\ee
The $D_{\mu}\psi(x, g_p)$'s are defined by Eqs. (2.4) and (2.5).
The $D_1$ and $D_2$ are covariant derivatives on $Z_3$ defined by
{
\setcounter{enumi}{\value{equation}}
\addtocounter{enumi}{1}
\setcounter{equation}{0}
\renewcommand{\theequation}{\arabic{section}.\theenumi\alph{equation}}
\bea
D_1 \psi(p)&=& \psi(p) - H(p, p+1) \psi(p+1),\\
D_2 \psi(p)&=& \psi(p)-H(p, p+2) \psi(p+2),
\eea
\setcounter{equation}{\value{enumi}}}
where $\psi(p) =\psi(x, g_p)$ and $H(p, p+i)$ is the mapping function
of $\psi(p+i)$ from $g_{p+i}$ to $g_p$.

Eqs. (3.3a) and (3.3b) are multiplied by $i \bar{\psi}(p) \gamma_5 \tau_1$
and $i \bar{\psi}(p) \gamma_5 \tau_2$, respectively, to yield
{
\setcounter{enumi}{\value{equation}}
\addtocounter{enumi}{1}
\setcounter{equation}{0}
\renewcommand{\theequation}{\arabic{section}.\theenumi\alph{equation}}
\bea
i\bar{\psi}(0) \gamma_5 \tau_1 D_1 \psi(0)
&=& -i \bar{\psi}(0) \gamma_5 \tau_1 H(0, 1) \psi(1)
= -\f{i}{\sqrt{2}}\bar{\psi}_5 \gamma_5 H(0, 1) \psi_{10},\nonumber\\
i\bar{\psi}(1) \gamma_5 \tau_1 D_1 \psi(1)
&=& -i \bar{\psi}(1) \gamma_5 \tau_1 H(1, 2) \psi(2)=0,\\
i\bar{\psi}(2) \gamma_5 \tau_1 D_1 \psi(2)
&=& -i \bar{\psi}(2) \gamma_5 \tau_1 H(2, 0) \psi(0)
= -\f{i}{\sqrt{2}}\bar{\psi}_{10} \gamma_5 H(2, 0) \psi_5,\nonumber
\eea
and
\bea
i\bar{\psi}(0) \gamma_5 \tau_2 D_2 \psi(0)
&=& -i \bar{\psi}(0) \gamma_5 \tau_2 H(0, 2) \psi(2)
= \f{1}{\sqrt{2}}\bar{\psi}_5 \gamma_5 H(0, 2) \psi_{10},\nonumber\\
i\bar{\psi}(1) \gamma_5 \tau_2 D_2 \psi(1)
&=& -i\bar{\psi}(1) \gamma_5 \tau_2 H(1, 0) \psi(0)
= -\f{1}{\sqrt{2}}\bar{\psi}_{10} \gamma_5 H(1, 0) \psi_5,\ \ \\
i\bar{\psi}(2) \gamma_5 \tau_2 D_2 \psi(2)
&=& -i \bar{\psi}(2) \gamma_5 \tau_2 H(2, 1) \psi(1)=0.\nonumber
\eea
\setcounter{equation}{\value{enumi}}
}
Thus, we have obtained
\bea
Y&\equiv& \sum_{p=0}^2 i \bar{\psi}(p) \gamma_5 \l(\tau_1 D_1 + \tau_2 D_2\r) \psi(p)\nonumber\\
&=& -\f{i}{\sqrt{2}} \bar{\psi}_{10} \gamma_5\l[H(2,0)-iH(1,0)\r]
-\f{i}{\sqrt{2}}\bar{\psi}_5 \gamma_5 \l[H(0, 1)+iH(0,2)\r] \psi_{10}.
\nonumber\\& &
\eea

The above equation is reduced to
\be
Y = -i \bar{\psi}_{10} \gamma_5 \f{1-i}{\sqrt{2}} H_5 \psi_5
-i \bar{\psi}_5 \gamma_5 \f{1+i}{\sqrt{2}} H_5^{\dagger} \psi_{10},
\ee
since one may set
\be
H(1,0)=H(2,0)=H_5(x),\ H(0,1)=H(0,2)=H_5^{\dagger}(x)
\ee
there.
The undesirable factors $i \gamma_5 (1\pm i)/\sqrt{2}$ can be written as
\be
i \gamma_5 = e^{i\pi \gamma_5/2},\ \f{1\pm i}{\sqrt{2}}=e^{\pm i\pi/4}.
\ee
Accordingly these factors are removed by redefining $\psi$'s and $H_5$ as
\be
e^{i\pi \gamma_5/4} \psi_5 \to \psi_5,\ e^{i\pi \gamma_5/4} \psi_{10} 
\to \psi_{10}
\ee
and
\be
e^{-i \pi/4} H_5 \to H_5,\ e^{i\pi/4} H_5^{\dagger} \to H_5^{\dagger}.
\ee

Thus the fermionic Lagrangian (3.1) becomes
\be
{\cal L}_F = i \bar{\psi}_5 \gamma^{\mu} D_{\mu} \psi_5 
+ i\bar{\psi}_{10} \gamma^{\mu} D_{\mu} \psi_{10}
- \kappa\l(\bar{\psi}_{10} H_5 \psi_5 + \bar{\psi}_5 H_5^{\dagger} \psi_{10}
\r),
\ee
where the $\kappa$-term is just the Yukawa coupling term, and $D_{\mu}\psi_5$
and $D_{\mu} \psi_{10}$ are defined by Eqs. (2.4) and (2.5), respectively.

The final Lagrangian for SU(5) GUT is, therefore, given by ${\cal L}_B$ in
Eq. (2.34) with $V(\phi, H) = V_f(\phi, H)$ given by (2.36)
and by ${\cal L}_F$ in Eq. (3.1), where the Higgs fields $\phi_{24}$
and $H_5$ should be rescaled as
\be
\sqrt{A} \phi_{24} = \Phi\ \mbox{and}\ \sqrt{B} H_5 = K.
\ee

\section{Concluding remarks}
\setcounter{equation}{0}

We have derived the SU(5) GUT from the geometrical view point of gauge 
theory on $M_4 \times Z_3$ manifold.
Compared with other derivations based on NCG, 
the geometric origin of SU(5) GUT has become clearer.

Contrary to the theory defined on $M_4 \times Z_2$,
which is the manifold of WS model, the derived Higgs 
potential (2.36) is not so rigidly fixed as it gives rise to
spontaneously broken gauge
symmetry. In order for symmetry breaking to occur one needs conditions
$a > 3c/8$ and $b > 4d$. In this case one can enjoy with a conjecture that 
the $Z_2$-type curvature coefficients $a$ and $b$ are well dominating over
the triangle-type curvature coefficients $c$ and $d$.

Here it is worthwhile to comment on the Yukawa coupling term $H^{\dagger}
\phi H$ appeared in our Higgs potential $V_f(\phi, H)$, see
Eqs. (2.36) and (2.37). In the phenomenological potential $V_c(\phi, H)$,
one drops such a term, including the term $\tr \phi^3$
as well, by imposing a discrete symmetry $\phi \to -\phi$.\cite{ref:8}
However, there are no compelling reasons in our approach to
introduce such a symmetry.
There are two possible ways to avoid such a term.
One is to impose some relations among parameters $c, d$
and $e$, i.e., to put $e=6(c+2d)$, and the other is to discard the
curvatures arising from the triangle graphs in Fig. 7 altogether.
Both methods are, however, not very convincing to the present authors. 

Finally let us briefly discuss a derivation of SO(10) GUT.\cite{ref:9}
As a typical model 
of SO(10) Mohapatra and Parida\cite{ref:10} considered three kinds of Higgs 
fields $\phi_{10},\ \phi_{126}$ and $\phi_{210}$
belonging to the 10-, 126-, and 210-dimensional representations, respectively,
of SO(10). In quite the same way as the SU(5) case this 
model can be interpreted also as geometrical gauge theory on $M_4 \times
Z_3$, where 
these three Higgs fields can be regarded as mapping functions of fermionic 
fields between three discrete points. As for the fermionic fields we may 
choose them to be the same representation given in Ref. 10 on each discrete 
points.

\appendix
\section{} %Empty argument \section{} yields `Appendix'. 
\setcounter{equation}{0}
\renewcommand{\theequation}{\thesection.\arabic{equation}}

We show that the mapping function (2.15a) cannot be unitary.
Denote the Hermitian conjugate of $H^i(x)$ by the complex conjugation
$*$ as
\be
\l(H^i(x)\r)^* \equiv H_i^{\dagger}(x),
\ee
then
\bea
\l({H^{ij}}_l(x, g_1, g_0)\r)^*&=&
\f{1}{2}\l[\l(H^i(x)\r)^*\delta^j_l-\l(H^j(x)\r)^*\delta^i_l\r]\nonumber\\
&=&\f{1}{2} \l[H^{\dagger}_i(x) \delta^l_j-H^{\dagger}_j(x)\delta^l_i\r]
\nonumber\\
&\equiv& {H_{ij}}^l(x, g_0, g_1).
\eea
The Hermitian conjugate of Eq. (2.14) is, therefore, given by
\be
\bar{\psi}_{\| ij}(x, g_1)=\bar{\psi}_l(x, g_0) {H_{ij}}^l(x, g_0, g_1).
\ee
The inverse mapping of Eq. (2.14) is defined in terms of this function
${H_{ij}}^l(x, g_0, g_1)$ as
{\setcounter{enumi}{\value{equation}}
\addtocounter{enumi}{1}
\setcounter{equation}{0}
\renewcommand{\theequation}{\thesection.\theenumi\alph{equation}}
\be
\psi_{\|}^l(x, g_0)={H_{ij}}^l(x, g_0, g_1) \psi_{\|}^{ij}(x, g_1)
\ee
and
\be
\bar{\psi}_{\| l}(x, g_0) = \bar{\psi}_{\| ij}(x, g_1)
{H^{ij}}_l(x, g_1, g_0),
\ee
\setcounter{equation}{\value{enumi}}}
$\!\!$where
$\psi_{\|}^l(x,g_0)\ (\bar{\psi}_{\| l}(x, g_0))$ is the image
of $\psi_{\|}^{ij}(x, g_1)\ (\bar{\psi}_{\| ij}(x, g_1))$ from 
$(x, g_1)$ to $(x, g_0)$. 

Substituting Eq. (2.14) into Eq. (A.4a) we have
\be
\psi_{\|}^l(x, g_0) = {H}_{ij}^{\ \ l}(x, g_0, g_1){H^{ij}}_k(x, g_1, g_0)
\psi^k(x, g_0).
\ee
The right hand side of Eq. (A.5) also appears in the norm
\be
\bar{\psi}_{\| ij}(g_1) \psi_{\|}^{ij}(g_1)
= \bar{\psi}_l(g_0) {H_{ij}}^l(g_0, g_1) {H^{ij}}_k(g_1, g_0) \psi^k(g_0).
\ee
Now, let us assume that the mapping function (2.15a) is unitary, i.e.,
\be
{H_{ij}}^l(x, g_0, g_1){H^{ij}}_k(x, g_1, g_0) = \delta^l_k.
\ee
Substituting Eqs. (2.15a) and (A.2) into Eq. (A.7), we find
\bea
\lefteqn{\f{1}{4}\l(H_i^{\dagger} \delta_j^l - H_j^{\dagger} \delta_i^l\r)
\l(H^i\delta^j_k-H^j \delta^i_k \r)}\nonumber\\
&=& \f{1}{2}\l[\l(H^{\dagger} H\r)\delta^l_k - H^{\dagger}_k H^l\r]
=\delta^l_k.
\eea
By taking the trace of the both sides, one obtains
\be
\l(H^{\dagger} H \r) = \f{5}{2}.
\ee
Substituting this into Eq. (A.8), we see for $l=k$
\be
H_l^{\dagger} H^l = \f{1}{2}\ \ \mbox{(no sum over $l$),}
\ee
and for $l \neq k$
\be
H_k^{\dagger} H^l = 0.
\ee

Last two equations (A.10) and (A.11) contradict each other.
This shows that the mapping function (2.15a) cannot be unitary,
and
\bea
\psi_{\|}^l(x, g_0) &\neq& \psi^l(x, g_0)\\
\bar{\psi}_{\| ij}(x, g_1) \psi_{\|}^{ij}(x, g_1) &\neq&
\bar{\psi}_l(x, g_0) \psi^l(x, g_0).
\eea


\begin{thebibliography}{99}
\bibitem{ref:1}
%
A.~Connes, in {\it The Interface of Mathematics and Particle Physics},
ed. D. Quilen, G.B. Segal and S.T. Tsou
(Clarendon Press，Oxford, 1990).\\
A.~Connes, {\it Noncommutative Geometry}, (Academic Press, London, 1994).\\
A.~Connes and J.~Lott，Nucl. Phys. Proc. Suppl. {\bf 18B} (1990) 29.\\
A.~Connes, J. Math. Phys. {\bf 36} (1995) 6194.\\
D.~Kastler, Rev. in Math. Phys. {\bf 5} (1990) 477.
%
\bibitem{ref:2}
%
R.~Coquereaux, G.~Esposito-Farese and G.~Vaillant, 
Nucl. Phys. {\bf B353} (1991) 689.\\
R.~Coquereaux, R.~H\"assling, N.A.~Papadopoulos and F.~Scheck,
Int. J. Mod. Phys. {\bf A7} (1992) 2809.\\
A.~Sitarz, Phys. Lett. {\bf B308} (1993) 311.\\
A.H.~Chamseddine, G.~Feldler and J.~Fr\"ohlich, Phys. Lett.
{\bf B296} (1992) 109;
Nucl. Phys. {\bf B395} (1993) 672.\\
K.~Morita, Prog. Theor. Phys. {\bf 90} (1993) 219.\\
H.G.~Ding, H.Y.~Guo, J.M.~Li and K. Wu, Z.~Phys. {\bf C64} (1994) 512.\\
K.~Morita and Y.~Okumura, Prog. Theor. Phys. {\bf 91}
(1994) 959; Phys. Rev. {\bf D50} (1994) 1016.\\
Y.~Okumura, Phys. Rev. {\bf D50} (1994) 1026; Prog. Theor. Phys.
{\bf 92} (1994) 625.\\
S.~Naka and E.~Umezawa, Prog. Theor. Phys. {\bf 92} (1994) 189.\\
R.~Erdem, Mod. Phys. Lett. {\bf A13} (1998) 465.
%
\bibitem{ref:3}
%
S. Weinberg, Phys. Rev. Lett. {\bf 19} (1967) 1264.\\
A. Salam, in {\it Elementary Particle Theory: Relativistic Group and 
Analyticity} (Nobel Symposium No.8), ed. N. Svartholm (Almqvist 
and Wiksell, Stockholm, 1968).
%
\bibitem{ref:4}
%
G. Konisi and T. Saito, Prog. Theor. Phys. {\bf 95} (1996) 657.
%
\bibitem{ref:5}
%
T. Saito and K. Uehara，Phys. Rev. {\bf D56} (1997) 2390.\\
B. Chen, T. Saito, H-B. Teng, K. Uehara and K. Wu, Prog. Theor. Phys.
{\bf 95} (1996) 1173.
A. Kokado, G. Konisi, T. Saito and Y. Tada, Prog. Theor. Phys.
{\bf 99} (1998) 293.
%
\bibitem{ref:6}
B. Pati and A. Salam, Phys. Rev. {\bf D8} (1973) 1240.\\
H. Georgi and S.L. Glashow, Phys. Rev. Lett. {\bf 32} (1974) 438.\\
H. Georgi, in {\it Particles and Fields} ed. C.E. Carlson, (AIP,
New York, 1975)\\
H. Fritzsch and P. Minkowski, Ann. Phys. {\bf 93} (1975) 193.
\bibitem{ref:7}
A.H. Chamseddine, G. Felder, and J. Fr\"ohlich, Phys. Lett. {\bf B296}
(1992) 109.\\
Y. Okumura, Phys. Rev. {\bf D50} (1994) 1026.\\
I.S. Sogami, Prog. Theor. Phys. {\bf 94} (1995) 117.\\
I.S. Sogami, Prog. Theor. Phys. {\bf 95} (1996) 637.
\bibitem{ref:8}
E. Gildener, Phys. Rev. {\bf D14} (1976) 1667.\\
A.J. Buras, J. Ellis, M.K. Gaillard and D.V. Nanopoulos, 
Nucl. Phys. {\bf B135} (1978) 66.
\bibitem{ref:9}
A.H. Chamseddine and J. Fr\"ohlich, Phys. Rev. {\bf D50} (1994) 2893.\\
Y. Okumura, Prog. Theor. Phys. {\bf 94} (1995) 607.
\bibitem{ref:10}
R.N. Mohapatra and M.K. Parida, Phys. Rev. {\bf D47} (1993) 264.\\
R.N. Mohapatra, {\it Unification and Supersymmetry} (Second Edition)，
(Springer-Verlag, New York, 1996)．
\end{thebibliography}
\end{document}